\documentclass[letterpaper,pdftex,rmp,twocolumn,amsmath,amssymb,groupedaddress,floatfix]{revtex4}
\RequirePackage[sort&compress]{natbib} %References in order as cited

\usepackage[utf8]{inputenc}
\usepackage[T1]{fontenc}
\usepackage{ae}
\usepackage{aecompl}

\usepackage[a]{esvect}

\usepackage{graphicx}% Include figure files
\usepackage{dcolumn}% Align table columns on decimal point
\usepackage{bm}% bold math

\usepackage{pgfplots}

\definecolor{darkgray}{rgb}{0.5,0.5,0.5}
\definecolor{darkred}{rgb}{0.89,0.10,0.11}
\definecolor{darkblue}{rgb}{0.12,0.39,0.62}
\usepackage{url}
\usepackage{enumitem}
\usepackage{tikz}
\usetikzlibrary{calc}

\usepackage[pdftex,breaklinks=true,colorlinks=true,citecolor=black,linkcolor=black,menucolor=black,urlcolor=darkblue,pdfborder={1 0 0}]{hyperref}
\hypersetup{pdftitle={Journal ranking by network flows with different amounts of memory},pdfauthor={L. Bohlin, A. V. Esquivel, A. Lancichinetti, and M. Rosvall (2014)}}
\bibpunct{(}{)}{,}{n}{,}{,}

\usepackage[font=sf,size=footnotesize,justification=RaggedRight,singlelinecheck=false]{caption}
\usepackage[sf,bf,small]{titlesec}
\titlespacing*{\section}{0pt}{1em}{0em}
\titlespacing*{\subsection}{0pt}{1em}{0em}
\titlespacing*{\subsubsection}{0pt}{1em}{0em}

\renewcommand\thesection{\arabic{section}}
\renewcommand\thesubsection{\thesection.\arabic{subsection}}

\makeatletter
\renewcommand{\p@section}{\arabic{section}\expandafter\@gobble}
\renewcommand{\p@subsection}{\thesection.\arabic{subsection}\expandafter\@gobble}
\renewcommand{\p@subsubsection}{\thesubsection.\arabic{subsubsection}\expandafter\@gobble}
\makeatother

\titleformat*{\section}{\large\bfseries\sffamily}
\titleformat*{\subsection}{\normalsize\bfseries\sffamily}
\titleformat*{\subsubsection}{\normalsize\sffamily}

\everymath=\expandafter{\the\everymath\displaystyle}

\begin{document}
\makeatletter
\renewcommand\@biblabel[1]{#1.}
\makeatother

%\preprint{APS/123-QED}

%%%%%%%%%%%%%%%%%%%%%%%%%%%%%%%%%%%%%%%%%%%%%%%%%%%%%%%%%%%%%%%%%%%%%%%%%%%%
%%%%%%%%%%%%%%%%%%%%%%%%%%%%%%%%%%%%%%%%%%%%%%%%%%%%%%%%%%%%%%%%%%%%%%%%%%%%
%%%%%%%%%%%%%%%%%%%%%%%%%%%%%%%%%%%%%%%%%%%%%%%%%%%%%%%%%%%%%%%%%%%%%%%%%%%%
%%%%%%%%%%%%%%%%%%%%%%%%%%%%%%%%%%%%%%%%%%%%%%%%%%%%%%%%%%%%%%%%%%%%%%%%%%%%
\title{Robustness of journal rankings by network flows\\ with different amounts of memory}% Force line breaks with \\

\author{Ludvig Bohlin}
\email{ludvig.bohlin@physics.umu.se}
\thanks{Corresponding author}
\author{Alcides Viamontes Esquivel}
\email{a.viamontes.esquivel@physics.umu.se}
\author{Andrea Lancichinetti}
\email{andrea.lancichinetti@physics.umu.se}
\author{Martin Rosvall}%
\email{martin.rosvall@physics.umu.se}
\affiliation{%
Integrated Science Lab, Department of Physics, Ume\r{a} University, SE-901 87 Ume\r{a}, Sweden\\
}%

\date{\today}% It is always \rightarrowday, today,

\begin{abstract}
As the number of scientific journals has multiplied, journal rankings have become increasingly important for scientific decisions. From submissions and subscriptions to grants and hirings, researchers, policy makers, and funding agencies make important decisions with influence from journal rankings such as the ISI journal impact factor. Typically, the rankings are derived from the citation network between a selection of journals and unavoidably depend on this selection. However, little is known about how robust rankings are to the selection of included journals. Here we compare the robustness of three journal rankings based on network flows induced on citation networks. They model pathways of researchers navigating scholarly literature, stepping between journals and remembering their previous steps to different degree: zero-step memory as impact factor, one-step memory as Eigenfactor, and two-step memory, corresponding to zero-, first-, and second-order Markov models of citation flow between journals. We conclude that higher-order Markov models perform better and are more robust to the selection of journals. Whereas our analysis indicates that higher-order models perform better, the performance gain for the second-order Markov model comes at the cost of requiring more citation data over a longer time period.

\end{abstract}

\pacs{Valid PACS appear here}% PACS, the Physics and Astronomy
\maketitle

%%%%%%%%%%%%%%%%%%%%%%%%%%%%%%%%%%%%%%%%%%%%%%%%%%%%%%%%%%%%%%%%%%%%%%%%%%%%
%%%%%%%%%%%%%%%%%%%%%%%%%%%%%%%%%%%%%%%%%%%%%%%%%%%%%%%%%%%%%%%%%%%%%%%%%%%%
%%%%%%%%%%%%%%%%%%%%%%%%%%%%%%%%%%%%%%%%%%%%%%%%%%%%%%%%%%%%%%%%%%%%%%%%%%%%
%%%%%%%%%%%%%%%%%%%%%%%%%%%%%%%%%%%%%%%%%%%%%%%%%%%%%%%%%%%%%%%%%%%%%%%%%%%%
%\section{Introduction}

Science builds on previous science in a recursive quest for new knowledge \cite{de1986little,pickering1992science,hull2010science}. Researchers put great effort into finding the best work by other researchers and into achieving maximum visibility of their own work. Therefore, they both search for good work and seek to publish in prominent journals. Inevitably, where researchers publish becomes a proxy for how good their work is, which in turn influences decisions regarding hiring, promotion, and tenure, as well as university rankings and academic funding \cite{harnad2004access,weingart2005impact}.  As a consequence, researchers depend on the perceived importance of the journals they publish in.
While actually reading the work published in a journal is the only way to qualitatively evaluate the scientific content, different metrics are nevertheless used to quantitatively assess the importance of scientific journals \cite{garfield1972citation,pinski1976citation,seglen1997impact,garfield1999journal,garfield2006history,plos2006impact,bollen2006journal,bergstrom2007measuring}.
In different ways, the metrics extract information from the network of citations between articles published in the journals.

In this paper, we analyze three flow-based journal rankings \cite{bollen2006journal,bergstrom2007measuring,rosvall2014memory} that at different order of approximations seek to capture the pathways of researchers navigating scholarly literature. Specifically, the metrics measure the journal visit frequency of random walk processes that correspond to zero-, first-, and second-order Markov models. That is, given a citation network between journals and a random walker following the citations, movements in a zero-order model are independent of the currently visited journal, movements in a first-order model depend only on the currently visited journal, and movements in a second-order model depend both on the currently visited journal and the previously visited journal.

Evaluating ranking methods inevitably becomes subjective, as their objectives often are different. Which method is best, the most transparent \cite{seglen1997impact}, the most difficult to game \cite{plos2006impact,braun2010improve}, or the one with highest predictive power \cite{stringer2008effectiveness,acuna2012future,penner2013predictability}? Irrespective of the specific objective, perhaps the most important criterion is nevertheless the robustness of the method \cite{vanclay2007robustness,ghoshal2011ranking}. Because journal rankings depend on the selection of journals included in the analysis, we compare the robustness of rankings obtained with zero-, first-, and second-order Markov models with random resampling techniques.

We first describe the commonly used metrics impact factor and Eigenfactor, which correspond to specific implementations of zero- and first-order Markov models, respectively. Then we put them in the same mathematical framework and show how a second-order Markov model can be devised in a similar way. We use data from Thomson Reuters Web of Science and compare the methods both qualitatively and quantitatively in terms of ranking order, ranking score distributions, and robustness.

\section*{Impact factor and Eigenfactor}

Impact factor was first described in 1972 and the ISI journal impact factor is today the leading indicator of journal influence \cite{garfield1972citation,garfield2006history}, despite its weaknesses \cite{seglen1997impact}. The impact factor of a journal in a given year measures the average number of citations to recent articles from articles published in the given year. The conventional two-year impact factor is calculated based on citation data from a three-year period. For example, the impact factor of journal $J$ in 2014 is the ratio between the number of citations from all considered journals in source year 2014 to articles published in $J$ in target years 2012--2013 and the number of articles published in $J$ in the target years. A five-year impact factor is calculated in a similar way with a five-year target window. The advantage with the widely used impact factor is that it is easy to calculate and explain, once the selection of journals is made.

Even though impact factor is not seen as a flow-based metric, it is in fact an example of a zero-order Markov model of flow between journals. The simple count of citations to journals corresponds to measuring the visit frequency of a random walker that visits journals proportional to their citation counts. While the measure is widely used, a major drawback is the underlying assumption that all citations carry equal weight, irrespective of origin.

Several rankings have been suggested to overcome the problem with uniform citation weights \cite{pinski1976citation,bollen2006journal,bergstrom2007measuring}. The Eigenfactor score \cite{bergstrom2007measuring, bergstrom2008eigenfactor} and its per-article normalized Article Influence Score, for example, builds on the PageRank algorithm \cite{brin1998anatomy} and takes advantage of the entire network of citations. Generally speaking, the Eigenfactor score measures the relative journal visit rate of a random walker that navigates between journals by following random citations. Therefore, the Eigenfactor score of a journal can be interpreted as a proxy for how often a researcher who randomly navigates the citation landscape  accesses content from the journal. In this way, the Eigenfactor score corresponds to a first-order Markov model for evaluating journal influence. In a recursive fashion, important journals are those that are highly cited by important journals. In practice, a citation from an influential journal will be worth more than a citation from a less significant journal, because its importance is inherited from the citing journal. However, the inherited importance is aggregated across a journal and pushed further no matter where it came from. As a result, the actual inheritance structure of the article-level citation network is lost with strongest effect on multidisciplinary journals.

While the main difference between impact factor and Eigenfactor is that they correspond to a zero- and a first-order Markov model of flow between journals, respectively, they differ in two other ways as well. First, while the conventional impact factor uses a two-year citation target window, Eigenfactor uses a five-year target window by default.
The extended time window was introduced because, in many fields, articles are not frequently cited until several years after publication.
Moreover, the Eigenfactor score considers inheritance of importance \emph{between} journals and therefore ignores self-citations. As a result, the incentive to boost the ranking of a journal with self-citations vanishes. In this paper, we focus on the general effects of Markov order rather than specific implementations. Therefore, we exclusively study rankings with five-year target windows and exclude all self-citations.

%%%%%%%%%%%%%%%%%%%%%%%%%%%%%%%%%%%%%%%%%%%%%%%%%%%%%%%%%%%%%%%%%%%%%%%%%%%%
%%%%%%%%%%%%%%%%%%%%%%%%%%%%%%%%%%%%%%%%%%%%%%%%%%%%%%%%%%%%%%%%%%%%%%%%%%%%
%%%%%%%%%%%%%%%%%%%%%%%%%%%%%%%%%%%%%%%%%%%%%%%%%%%%%%%%%%%%%%%%%%%%%%%%%%%%
%%%%%%%%%%%%%%%%%%%%%%%%%%%%%%%%%%%%%%%%%%%%%%%%%%%%%%%%%%%%%%%%%%%%%%%%%%%%

\section*{Modeling citation flow}

To model citation flows between journals, we first aggregate article-level citation data in journals and then model the network flow with a random walk process. We construct citation flows with different amounts of memory by aggregating the citation data in networks that correspond to zero-, first-, and second-order Markov models. Below we in turn describe how we aggregate the data and model the flow. 

\subsection*{Journal networks with different amounts of memory}

\begin{figure*}[tbp]
\includegraphics[width=\textwidth]{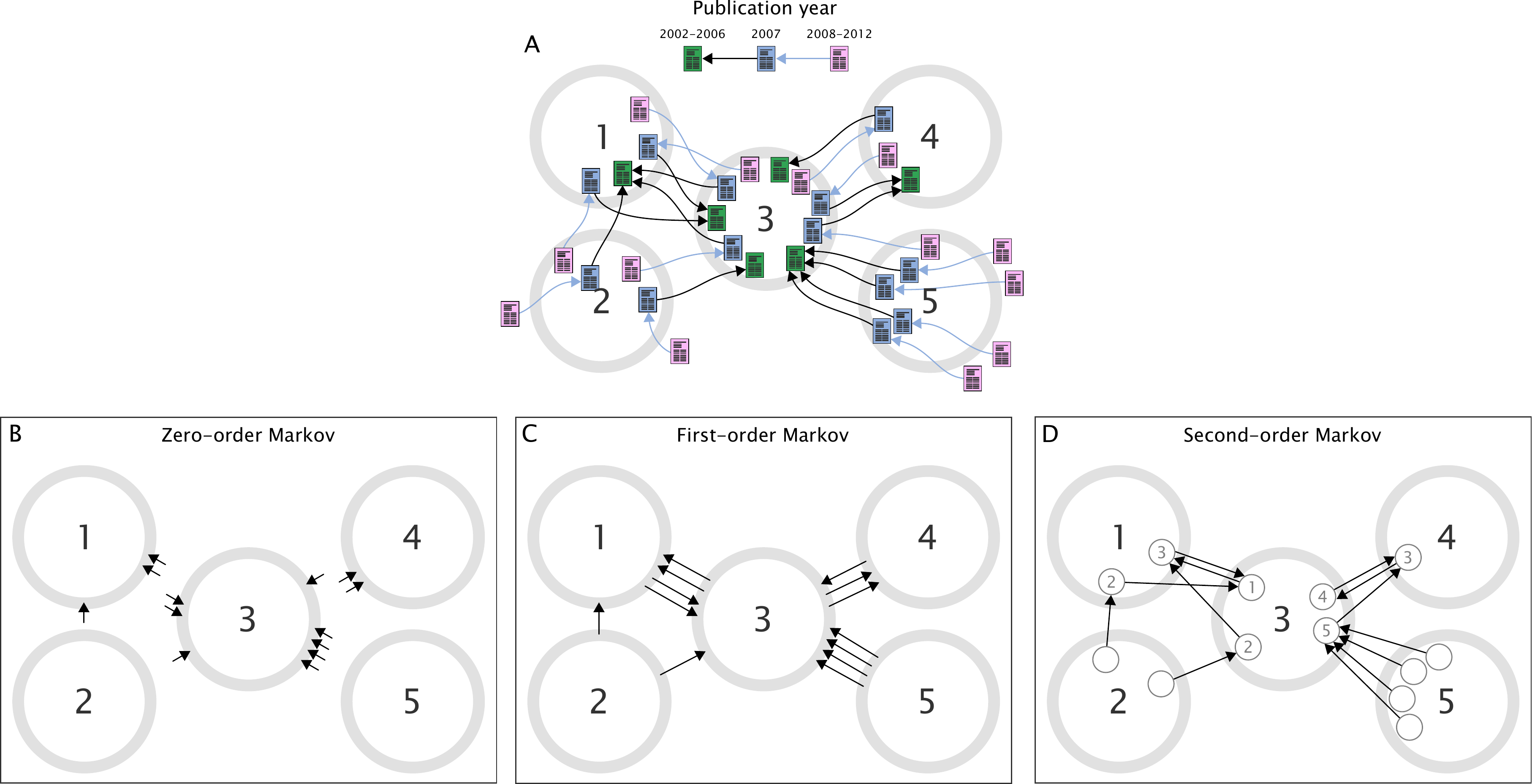}
\caption{From an article-level citation network to journal-level citation networks with different amount of memory. (\textsf{A}) A citation network with articles published in the early source year 2007 (blue), cited by articles published in the late source years 2008--2012 (pink), and citing articles published in early target years 2002--2006 (green). Gray circles represent journals. (\textsf{B}) A zero-order Markov model defines movements that only depend on the number of incoming citations of a journal. For clarity, only incoming links are shown. (\textsf{C}) A first-order Markov model defines movements that depend on the number of incoming citations of a journal from the currently visited journal. (\textsf{D}) A second-order Markov model defines movements between memory nodes, such that movements between journals depend on the currently visited journal and the previously visited journal. \label{fig:schematic}}
\end{figure*}

We use article-level citation data from Thomson Reuters Web of Science 1980-2013. The data include almost one billion citations between more than 30 million articles published in about 20,000 journals. In this study, we focus on articles published in the years 2007--2012 and their citations to articles published in 2002-2007.  Specifically, we are interested in articles published in 2007, their citations to articles published in 2002--2006, and citations to the articles published in 2007 from articles published in 2008--2012. We need the two overlapping time windows to construct the second-order Markov model.

Figure \ref{fig:schematic} illustrates how we construct journal citation networks with different amount of memory from article-level citation data. In Fig.~\ref{fig:schematic}A, we show a schematic citation network with articles published in 11 different journals. The articles were published in three different time periods, the early target years 2002--2007, the early source year 2007, which also is the target year of the late source years 2008--2012. For the zero- and first-order Markov models, we used the early target and source years 2002--2007, and for the second-order Markov model we also included the late source years 2008--2012. We excluded proceedings, but included all journals $k=1,2,\ldots,N$ that received citations during the target period.

For the zero-order Markov model, we counted the number of citations to articles published in the early target years 2002-2006 from articles published in the early source year 2007. To construct the journal network, we aggregated these citations in the journals of the cited articles. That is, each citation $j \rightarrow k$ between an article published in journal $j$ in the early source year to an article published in journal $k$ in the early target years, adds a weight of one to the cited journal $k$, $W(k) \longrightarrow W(k) + 1$.

This procedure is exemplified in Figs.~\ref{fig:schematic}A and B, with articles published in the early target years in green and articles published in the early source year in blue. Figure \ref{fig:schematic}A shows how one article published in journal 1 receives three citations, how four articles published in journal 3 receive eight citations, and how one article published in journal 4 receives two citations. For this zero-order Markov network shown in Fig.~\ref{fig:schematic}B, journals are connected to other journals with weights proportional to the number of incoming citations, independent of citation source. That is, a random walk process on a zero-order Markov network is memoryless such that the next step does not depend on the currently visited journal.

For the first-order Markov model, we aggregated the citations described above in pairs of citing and cited journals. That is, each citation between an article published in journal $j$ in the early source year to an article published in journal $k$ in the early target years adds a link weight of one between the citing and the cited journals, $W(j \rightarrow k) \longrightarrow W(j \rightarrow k) + 1$. Figure \ref{fig:schematic}C illustrates how the 13 incoming links in the zero-order Markov model have specific sources of the citing journals in the first-order Markov model. Accordingly, a random walk process on a first-order Markov network has a one-step memory such that the next step depends on the currently visited journal.

For the second-order Markov model, we also included citations from articles published in the late source years. We used citation chains $i \rightarrow j \rightarrow k$, trigrams of articles published in journal $i$ in the late source years that cite articles in journal $j$ in the early source year that in turn cite articles in journal $k$ in the early target years, as illustrated in Fig.~\ref{fig:schematic}A. To construct the second-order Markov network, we aggregated the trigrams in memory nodes $\vv{ij}$, such that each citation chain $i \rightarrow j \rightarrow k$ adds a link weight of one between memory nodes $\vv{ij}$ and $\vv{jk}$, $W(\vv{ij} \rightarrow \vv{jk}) \longrightarrow W(\vv{ij} \rightarrow \vv{jk}) + 1$. That is, each journal has $n_j$ memory nodes, one for each other journal that cites it. Constructed in this way, a random walk process on a second-order Markov network has a two-step memory such that the next step depends not only on the currently visited journal, but also on the previously visited journal.

\begin{figure*}[tbp]
\includegraphics[width=\textwidth]{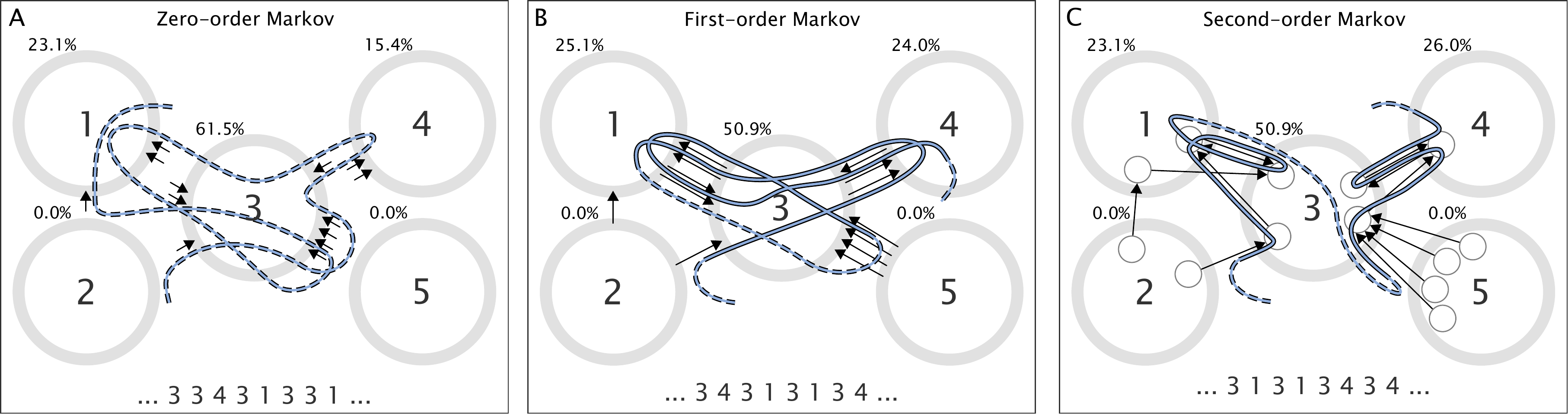}
\caption{Flow-based journal ranking based on zero-, first-, and second-order Markov models of citation flow. 
(\textsf{A}) In a zero-order Markov model, where a random walker's next move only depends on the number of incoming citations of a journal, and journal 1 receives significantly more flow than journal 4, for example. (\textsf{B}) In a first-order Markov model, where a random walker moves next depends on the number of incoming citations of a journal from the currently visited journal, and journal 4 receives almost as much flow as journal 1. (\textsf{C}) In a second-order Markov model, where a random walker moves next depends on the number of incoming citations of a journal from the currently visited journal and depending on the previously visited journal, and journal 4 receives more flow than journal 1. To make the flow independent of the starting position, at a low rate the random walker teleports to a journal proportional to its number of incoming citations (dashed trail). A journal's ranking is set by the relative frequency at which the random walker visits the journal.
\label{fig:schematicwalk}}
\end{figure*}

The procedure to construct a second-order Markov network above assumes that each article in the early source years only is cited by one journal in the later source years. For each citation from an article in journal $j$ in the early source year to an article in journal $k$ in the early target years $j \to k$, we identify all $n$ articles published in any journal $i$ in the late source years that cite the article in the early source year, and add a fractional link weight of $1/n$ between memory nodes $\vv{ij}$ and $\vv{jk}$. Moreover, if we cannot identify a trigram $i \rightarrow j \rightarrow k$, because the article in the early source year was never cited by an article in the late source years, we add a fractional link weight $1/n_j$ between memory nodes $\vv{ij}$ and $\vv{jk}$ for all $n_j$ memory nodes $\vv{ij}$ of journal $j$. In this way, we obtain the first-order Markov network if we aggregate the memory nodes in their respective journals.

\subsection*{Modeling citation flow with a random walker}

We use a random walk process on the networks with different amounts of memory to obtain the journal rankings.
The random walk processes can be seen as proxies for how researchers navigate scholarly literature, as they read articles and follow citations in their search for information. In the zero-order Markov model, a researcher would pick any citation and follow it to the cited journal irrespective of where the currently read article is published (Fig.~\ref{fig:schematicwalk}A). In the first-order Markov model, a researcher would pick a citation from any article published in the same journal as the currently read article and follow it to the cited journal (Fig.~\ref{fig:schematicwalk}B). In the second-order Markov model, a researcher would pick a citation from an article published in the same journal as the currently read article that is also cited by an article published in the previously visited journal and follow it to the cited journal (Fig.~\ref{fig:schematicwalk}C). In this way, the random walk processes correspond to researchers with zero-, one-, and two-step memory.

The first-, second-, and third order Markov models are obtained with the same random walk process on the three networks with zero-, one-, and two-step memory. Formally, we represent the journal visited at time $t$ by the random variable $X_t$. The random walk process generates a sequence of visited journals $X_{1} X_{2} \ldots X_{t}$. In general, the journal visited at time $t+1$ depends on the full history of the dynamic process,
\begin{align}
P(k;t+1) &\equiv P(X_{t+1} = k_{t+1}) \\
		%&= P(X_{t+1} = k_{t+1} | X_{t} = k_{t}, X_{t-1} = k_{t-1}, \ldots, X_{1} = k_{1} ), \nonumber
		&= P(X_{t+1} = k_{t+1} | X_{t} = k_{t}, \ldots, X_{1} = k_{1} ), \nonumber
\end{align}
but for the processes we consider here the memory is limited.

For the zero-order Markov model illustrated in Fig.~\ref{fig:schematicwalk}A, the probability to step to journal $k$ next is simply given by the relative number of citations to that journal irrespective of the currently visited journal,
\begin{align}
p(k) = \frac{W(k)}{\sum_{k} W(k)},\label{eq:zerotransition}
\end{align}
which therefore also is the stationary solution of the zero-order Markov model,
\begin{align}
\pi^{(0)}(k) = \frac{W(k)}{\sum_{k} W(k)}.\label{eq:zerosolution}
\end{align}

For the first-order Markov model illustrated in Fig.~\ref{fig:schematicwalk}B, the probability to step to journal $k$ next from journal $j$ is given by the relative number of citations to $k$ from $j$,
\begin{align}
p(j \rightarrow k) = \frac{W(j \rightarrow k)}{\sum_{k} W(j \rightarrow k)}.\label{eq:firsttransition}
\end{align}
Accordingly, the probability that the random walker visits node $k$ in step $t+1$ is in principle
\begin{align}
p(k;t+1) = \sum_j P(j;t)p(j \rightarrow k).
\end{align}
However, to ensure a unique solution independent of where the random walker is initiated, at a low rate $1-\alpha$ the random walker instead moves according to the zero-order Markov model,
\begin{align}
P(k;t+1) = \alpha \sum_j P(j;t) p(j \rightarrow k) + (1-\alpha)p(k),
\end{align}
with stationary solution given by
\begin{align}
\pi^{(1)}(k) = \alpha \sum_j \pi^{(1)}(j) p(j \rightarrow k) + (1-\alpha)p(k). \label{eq:firststationary}
\end{align}
The zero-order Markov step corresponds to random teleportation to journals proportional to their number of incoming citations. This link-weighted teleportation gives results that are more robust to changes in the teleportation rate $1-\alpha$ \cite{lambiotte2012ranking}. We use teleportation rate $1-\alpha = 0.15$ in all analyses. Note that this teleportation scheme is slightly different from the one used in Eigenfactor \cite{bergstrom2008eigenfactor}. However, unrecorded teleportation to a journal proportional to the number of articles it publishes followed by a recorded first-order Markov step, as used in Eigenfactor, is approximately the same as a single recorded zero-order Markov step. For example, they would be identical if all articles cited the same number of articles.

For the second-order Markov model illustrated in Fig.~\ref{fig:schematicwalk}C, the random walker moves from memory node to memory node proportional to the link weights between the memory nodes. For example, the probability to visit memory node $\vv{jk}$ after visiting memory node $\vv{ij}$ is
\begin{align}
p(\vv{ij} \rightarrow \vv{jk}) = \frac{W(\vv{ij} \rightarrow \vv{jk})}{\sum_k W(\vv{ij} \rightarrow \vv{jk})}.\label{eq:secondorderNextstep}
\end{align}
Accordingly, the probability that the random walker visits memory node $\vv{jk}$ in step $t+1$ is in principle
\begin{align}
\label{general1}
P(\vv{jk};t+1) = \sum_{i} P(\vv{ij};t) p(\vv{ij} \rightarrow \vv{jk}), 
\end{align}
but to ensure a unique solution we include teleportation steps also in this process,
\begin{align} 
P(\vv{jk};t+1) = \alpha \sum_{i} P(\vv{ij};t) p(\vv{ij} \rightarrow \vv{jk}) + (1-\alpha) p(\vv{jk}).
\end{align}
Here $p(\vv{jk})$ is given by the relative number of links to memory node $\vv{jk}$, which is equivalent to the relative number of links between node $j$ and $k$,
\begin{align}
p(\vv{jk}) = \frac{\sum_iW(\vv{ij} \rightarrow \vv{jk})}{\sum_{ijk}W(\vv{ij} \rightarrow \vv{jk})} = \frac{W(j \rightarrow k)}{\sum_{jk} W(j \rightarrow k)}.\label{eq:secondorder}
\end{align}
Consequently, the stationary solution is given by
\begin{align} 
\pi^{(2)}(\vv{jk}) = \alpha \sum_{i} \pi^{(2)}(\vv{ij}) p(\vv{ij} \rightarrow \vv{jk}) + (1-\alpha) p(\vv{jk}).\label{eq:secondstationary}
\end{align}
This teleportation scheme gives unbiased comparisons because journals receive the same amount of teleported flow as in the first-order Markov model, 
\begin{align}
\frac{\sum_{j} W(j \rightarrow k)}{\sum_{jk} W(j \rightarrow k)} = p(k),
\end{align}
and proportional to the stationary solution of the zero-order Markov model in Eq.~(\ref{eq:zerosolution}).

We obtain the nontrivial stationary solutions of Eq.~(\ref{eq:firststationary}) and (\ref{eq:secondstationary}) with the power-iteration method \cite{lanczos1950iteration}. For per-article rankings, analogous to the impact factor and the Article Influence Score, we simply divide the stationary solution of a journal by the number of articles published by that journal in the early target years. For easy comparison between the rankings of the different Markov models, we normalize with respect to the average journal. In this way, a ranking score of a journal larger than one tells how many times higher the stationary distribution per article is compared with the average journal. 

The common framework for the three ranking models makes it easy to study effects of the Markov order alone. However, the common framework also means that the models studied here are not identical to the established impact factor and Eigenfactor, and conclusions should be treated with care even if the differences are small. In summary, unlike impact factor, we disregard all self-links, and unlike Eigenfactor, we use recorded teleportation to journals proportional to their citation counts.

%%%%%%%%%%%%%%%%%%%%%%%%%%%%%%%%%%%%%%%%%%%%%%%%%%%%%%%%%%%%%%%%%%%%%%%%%%%%
%%%%%%%%%%%%%%%%%%%%%%%%%%%%%%%%%%%%%%%%%%%%%%%%%%%%%%%%%%%%%%%%%%%%%%%%%%%%
%%%%%%%%%%%%%%%%%%%%%%%%%%%%%%%%%%%%%%%%%%%%%%%%%%%%%%%%%%%%%%%%%%%%%%%%%%%%
%%%%%%%%%%%%%%%%%%%%%%%%%%%%%%%%%%%%%%%%%%%%%%%%%%%%%%%%%%%%%%%%%%%%%%%%%%%%

\section*{Results and discussion} %Flow-based journal rankings with different amount of memory
In this section, we show the results of comparisons between ranking scores obtained with zero-, first-, and second-order Markov models. We first show comparisons of the top journals in explicit ranking lists, and then show quantitative results for ranking scores and robustness.

\subsection*{Ranking scores}
Figure~\ref{fig:ranking} shows the rankings of the top 20 journals obtained with the three different Markov models. The ranking scores are given by the per article stationary distribution of random walkers normalized such that the average journal has score 1, as described above.
As with impact factor, review journals with few highly cited reviews have the highest rankings in all three models. They are followed by high impact multidisciplinary journals. Journals that lose from the zero- to the first-order flow model also tend to lose from the first- to the second-order model, and, vice versa, journals that gain from the zero- to the first-order flow model also tend to gain from the first- to the second-order model. However, the multidisciplinary journals only gain marginally from the first- to the second-order model. For the similar ranking analysis with the less complete and more biased citation data from JSTOR reported in ref.~\cite{rosvall2014memory}, the effect on multidisciplinary journals was even stronger because leaking flow between fields did not cancel to the same degree. In any case, and as schematically illustrated in Fig.~\ref{fig:schematicwalk}C, the relative rankings show the largest change from the zero- to the first-order model.

%=====================================================================
\newlength\x
\settowidth{\x}{99.9 Annu Rev Astron Astr}
\definecolor{paperred}{RGB}{163,49,84}
\definecolor{papergreen}{RGB}{49,163,84}
\definecolor{paperblue}{RGB}{84,49,163}
\definecolor{papergray}{RGB}{127,127,127}

\newcommand\tikzmark[1]{%
\tikz[remember picture,overlay]\node (#1) {};%
}

\newcommand\Connectsame[3][]{%
\tikz[remember picture,overlay]
\draw[->,black,>=latex,#1] ( $ (#2.north east) + (-3pt,-0.1em) $ ) -- ( $ (#3.north east) + (-20pt,-0.1em) $ );%
}

\newcommand\Connectincrease[3][]{%
tikz[remember picture,overlay]
\draw[->,paperblue,>=latex,#1] ( $ (#2.north east) + (-3pt,-0.1em) $ ) -- ( $ (#3.north east) + (-20pt,-0.1em) $ );%
}

\newcommand\Connectdecrease[3][]{%
\tikz[remember picture,overlay]
\draw[->,paperred,>=latex,#1] ( $ (#2.north east) + (-3pt,-0.1em) $ ) -- ( $ (#3.north east) + (-20pt,-0.1em) $ );%
}

\newcommand\Connectremoved[3][]{%
\tikz[remember picture,overlay]
\draw[->,paperred,dashed,>=latex,#1] ( $ (#2.north east) + (-3pt,-0.1em) $ ) -- ( $ (#3.north east) + (-20pt,-0.1em) $ );%
}

\newcommand\Connectnew[3][]{%
\tikz[remember picture,overlay]
\draw[->,paperblue,dashed,>=latex,#1] ( $ (#2.north east) + (-3pt,-0.1em) $ ) -- ( $ (#3.north east) + (-20pt,-0.1em) $ );%
}

\begin{figure*}

{\sffamily\footnotesize
\noindent\begin{minipage}[t]{0.33\textwidth}
\begin{enumerate}[label=\arabic*.,itemsep=0em]
\item[] {\bfseries Zero-order Markov \hfill \hfill}
\item \parbox{0.9\x}{{\color{papergray}34.6} Annu Rev Immunol \color{papergray}\dotfill} \tikzmark{start_same1}
\item \parbox{0.9\x}{{\color{papergray}27.8} Rev Mod Phys \color{papergray}\dotfill} \tikzmark{start1}
\item \parbox{0.9\x}{{\color{papergray}25.8} Ca-Cancer J Clin \color{papergray}\dotfill} \tikzmark{start2}
\item \parbox{0.9\x}{{\color{papergray}25.5} Physiol Rev \color{papergray}\dotfill} \tikzmark{start3}
\item \parbox{0.9\x}{{\color{papergray}24.4} Nat Rev Cancer \color{papergray}\dotfill} \tikzmark{start4}
\item \parbox{0.9\x}{{\color{papergray}23.7} New Engl J Med \color{papergray}\dotfill} \tikzmark{start5}
\item \parbox{0.9\x}{{\color{papergray}23.2} Annu Rev Biochem \color{papergray}\dotfill} \tikzmark{start6}
\item \parbox{0.9\x}{{\color{papergray}22.0} Nat Rev Immunol \color{papergray}\dotfill} \tikzmark{start_same2}
\item \parbox{0.9\x}{{\color{papergray}21.1} Annu Rev Neurosci \color{papergray}\dotfill} \tikzmark{start7}
\item \parbox{0.9\x}{{\color{papergray}20.4} Nat Rev Mol Cell Bio \color{papergray}\dotfill} \tikzmark{start8}
\item \parbox{0.9\x}{{\color{papergray}18.4} Chem Rev \color{papergray}\dotfill} \tikzmark{start_removed1}
\item \parbox{0.9\x}{{\color{papergray}18.1} Cell \color{papergray}\dotfill} \tikzmark{start9}
\item \parbox{0.9\x}{{\color{papergray}17.7} Annu Rev Cell Dev Bi \color{papergray}\dotfill} \tikzmark{start10}
\item \parbox{0.9\x}{{\color{papergray}17.3} Nat Med \color{papergray}\dotfill} \tikzmark{start11}
\item \parbox{0.9\x}{{\color{papergray}17.3} Nat Immunol \color{papergray}\dotfill} \tikzmark{start12}
\item \parbox{0.9\x}{{\color{papergray}17.2} Nature \color{papergray}\dotfill} \tikzmark{start13}
\item \parbox{0.9\x}{{\color{papergray}17.1} Science \color{papergray}\dotfill} \tikzmark{start14}
\item \parbox{0.9\x}{{\color{papergray}16.7} Nat Rev Neurosci \color{papergray}\dotfill} \tikzmark{start15}
\item \parbox{0.9\x}{{\color{papergray}16.3} Endocr Rev \color{papergray}\dotfill} \tikzmark{start_removed2}
\item \parbox{0.9\x}{{\color{papergray}15.5} Annu Rev Astron Astr \color{papergray}\dotfill} \tikzmark{start16}
\item[] \parbox{0.9\x}{\hfill \hfill} \tikzmark{zero_to_first_new}
\end{enumerate}
\end{minipage}%
\begin{minipage}[t]{0.33\textwidth}
\begin{enumerate}[label=\arabic*.,itemsep=0em]
\item[] {\bfseries First-order Markov \hfill \hfill} 
\item \tikzmark{start_same1end} \parbox{0.9\x}{{\color{papergray}54.0} Annu Rev Immunol \color{papergray}\dotfill}\tikzmark{start_same3}
\item \tikzmark{start6end} \parbox{0.9\x}{{\color{papergray}40.3} Annu Rev Biochem \color{papergray}\dotfill}\tikzmark{start_same4}
\item \tikzmark{start8end} \parbox{0.9\x}{{\color{papergray}35.2} Nat Rev Mol Cell Bio \color{papergray}\dotfill}\tikzmark{start17}
\item \tikzmark{start9end} \parbox{0.9\x}{{\color{papergray}33.9} Cell \color{papergray}\dotfill}\tikzmark{start18}
\item \tikzmark{start7end} \parbox{0.9\x}{{\color{papergray}33.7} Annu Rev Neurosci \color{papergray}\dotfill}\tikzmark{start19}
\item \tikzmark{start10end} \parbox{0.9\x}{{\color{papergray}33.1} Annu Rev Cell Dev Bi \color{papergray}\dotfill}\tikzmark{start20}
\item \tikzmark{start4end} \parbox{0.9\x}{{\color{papergray}33.0} Nat Rev Cancer \color{papergray}\dotfill}\tikzmark{start21}
\item \tikzmark{start_same2end} \parbox{0.9\x}{{\color{papergray}32.6} Nat Rev Immunol \color{papergray}\dotfill}\tikzmark{start22}
\item \tikzmark{start1end} \parbox{0.9\x}{{\color{papergray}32.4} Rev Mod Phys \color{papergray}\dotfill}\tikzmark{start23}
\item \tikzmark{start3end} \parbox{0.9\x}{{\color{papergray}29.6} Physiol Rev \color{papergray}\dotfill}\tikzmark{start24}
\item \tikzmark{start12end} \parbox{0.9\x}{{\color{papergray}29.3} Nat Immunol \color{papergray}\dotfill}\tikzmark{start25}
\item \tikzmark{start2end} \parbox{0.9\x}{{\color{papergray}26.4} Ca-Cancer J Clin \color{papergray}\dotfill}\tikzmark{start26}
\item \tikzmark{start5end} \parbox{0.9\x}{{\color{papergray}25.8} New Engl J Med \color{papergray}\dotfill}\tikzmark{start27}
\item \tikzmark{start13end} \parbox{0.9\x}{{\color{papergray}25.5} Nature \color{papergray}\dotfill}\tikzmark{start28}
\item \tikzmark{start_new1} \parbox{0.9\x}{{\color{papergray}24.4} Nat Genet \color{papergray}\dotfill}\tikzmark{start29}
\item \tikzmark{start14end} \parbox{0.9\x}{{\color{papergray}24.4} Science \color{papergray}\dotfill}\tikzmark{start_same5}
\item \tikzmark{start15end} \parbox{0.9\x}{{\color{papergray}23.4} Nat Rev Neurosci \color{papergray}\dotfill}\tikzmark{start30}
\item \tikzmark{start11end} \parbox{0.9\x}{{\color{papergray}22.3} Nat Med \color{papergray}\dotfill}\tikzmark{start_removed3}
\item \tikzmark{start16end} \parbox{0.9\x}{{\color{papergray}22.3} Annu Rev Astron Astr \color{papergray}\dotfill}\tikzmark{start_removed4}
\item \tikzmark{start_new2} \parbox{0.9\x}{{\color{papergray}21.9} Annu Rev Genet \color{papergray}\dotfill}\tikzmark{start31}
\item[] \tikzmark{first_removed_and_new1} \parbox{0.9\x}{\hfill \hfill} \tikzmark{first_removed_and_new2}
\end{enumerate}
\end{minipage}
\begin{minipage}[t]{0.33\textwidth}
\begin{enumerate}[label=\arabic*.,itemsep=0em]
\item[] {\bfseries Second-order Markov \hfill \hfill}
\item \tikzmark{start_same3end} \parbox{0.9\x}{{\color{papergray}56.3}  Annu Rev Immunol  \hfill \hfill}
\item \tikzmark{start_same4end} \parbox{0.9\x}{{\color{papergray}44.6}  Annu Rev Biochem  \hfill \hfill}
\item \tikzmark{start18end} \parbox{0.9\x}{{\color{papergray}39.1}  Cell  \hfill \hfill}
\item \tikzmark{start17end} \parbox{0.9\x}{{\color{papergray}39.0}  Nat Rev Mol Cell Bio  \hfill \hfill}
\item \tikzmark{start20end} \parbox{0.9\x}{{\color{papergray}38.0}  Annu Rev Cell Dev Bi  \hfill \hfill}
\item \tikzmark{start23end} \parbox{0.9\x}{{\color{papergray}36.7}  Rev Mod Phys  \hfill \hfill}
\item \tikzmark{start19end} \parbox{0.9\x}{{\color{papergray}36.4}  Annu Rev Neurosci  \hfill \hfill}
\item \tikzmark{start21end} \parbox{0.9\x}{{\color{papergray}33.5}  Nat Rev Cancer  \hfill \hfill}
\item \tikzmark{start22end} \parbox{0.9\x}{{\color{papergray}33.3}  Nat Rev Immunol  \hfill \hfill}
\item \tikzmark{start25end} \parbox{0.9\x}{{\color{papergray}32.0}  Nat Immunol  \hfill \hfill}
\item \tikzmark{start24end} \parbox{0.9\x}{{\color{papergray}28.3}  Physiol Rev  \hfill \hfill}
\item \tikzmark{start28end} \parbox{0.9\x}{{\color{papergray}27.6}  Nature  \hfill \hfill}
\item \tikzmark{start29end} \parbox{0.9\x}{{\color{papergray}27.1}  Nat Genet  \hfill \hfill}
\item \tikzmark{start26end} \parbox{0.9\x}{{\color{papergray}26.8}  Ca-Cancer J Clin  \hfill \hfill}
\item \tikzmark{start27end} \parbox{0.9\x}{{\color{papergray}26.6}  New Engl J Med  \hfill \hfill}
\item \tikzmark{start_same5end} \parbox{0.9\x}{{\color{papergray}25.9}  Science  \hfill \hfill}
\item \tikzmark{start_new3} \parbox{0.9\x}{{\color{papergray}25.0}  Nat Cell Biol  \hfill \hfill}
\item \tikzmark{start31end} \parbox{0.9\x}{{\color{papergray}24.1}  Annu Rev Genet  \hfill \hfill}
\item \tikzmark{start30end} \parbox{0.9\x}{{\color{papergray}23.6}  Nat Rev Neurosci  \hfill \hfill}
\item \tikzmark{start_new4} \parbox{0.9\x}{{\color{papergray}23.2}  Immunity  \hfill \hfill}
\item[] \tikzmark{first_to_second_removed}
\end{enumerate}
\end{minipage}
}
\centering
\begin{tikzpicture}[remember picture,overlay]
\Connectincrease{start8}{start8end}
\Connectincrease{start9}{start9end}
\Connectdecrease{start2}{start2end}
\Connectdecrease{start3}{start3end}
\Connectdecrease{start1}{start1end}
\Connectincrease{start6}{start6end}
\Connectincrease{start7}{start7end}
\Connectdecrease{start4}{start4end}
\Connectdecrease{start5}{start5end}
\Connectincrease{start29}{start29end}
\Connectincrease{start28}{start28end}
\Connectincrease{start25}{start25end}
\Connectdecrease{start24}{start24end}
\Connectdecrease{start27}{start27end}
\Connectdecrease{start26}{start26end}
\Connectdecrease{start21}{start21end}
\Connectincrease{start20}{start20end}
\Connectincrease{start23}{start23end}
\Connectdecrease{start22}{start22end}
\Connectsame{start_same4}{start_same4end}
\Connectsame{start_same5}{start_same5end}
\Connectsame{start_same1}{start_same1end}
\Connectsame{start_same2}{start_same2end}
\Connectsame{start_same3}{start_same3end}
\Connectincrease{start10}{start10end}
\Connectdecrease{start11}{start11end}
\Connectincrease{start12}{start12end}
\Connectincrease{start13}{start13end}
\Connectincrease{start14}{start14end}
\Connectincrease{start15}{start15end}
\Connectincrease{start16}{start16end}
\Connectdecrease{start17}{start17end}
\Connectincrease{start18}{start18end}
\Connectdecrease{start19}{start19end}
\Connectdecrease{start30}{start30end}
\Connectincrease{start31}{start31end}
\Connectremoved{start_removed1}{first_removed_and_new1}
\Connectremoved{start_removed2}{first_removed_and_new1}
\Connectremoved{start_removed3}{first_to_second_removed}
\Connectremoved{start_removed4}{first_to_second_removed}
\Connectnew{zero_to_first_new}{start_new2}
\Connectnew{zero_to_first_new}{start_new1}
\Connectnew{first_removed_and_new2}{start_new4}
\Connectnew{first_removed_and_new2}{start_new3}
\end{tikzpicture}
\caption{\label{fig:ranking}Gainers and losers among top journals. Comparison of journal rankings for zero-, first- and second-order Markov models of citation flow.
The ranking lists show the top 20 journals in 2007 for each model with citation data from Thomson Reuters Web of Science. Arrows connect journals from lower- to higher-order Markov models. Blue arrows for gainers, red arrows for losers, and black arrows for journals that do not change the rank order. Dashed arrows for journals that are not in the top 20 in all rankings. The ranking scores in gray.}
\end{figure*}
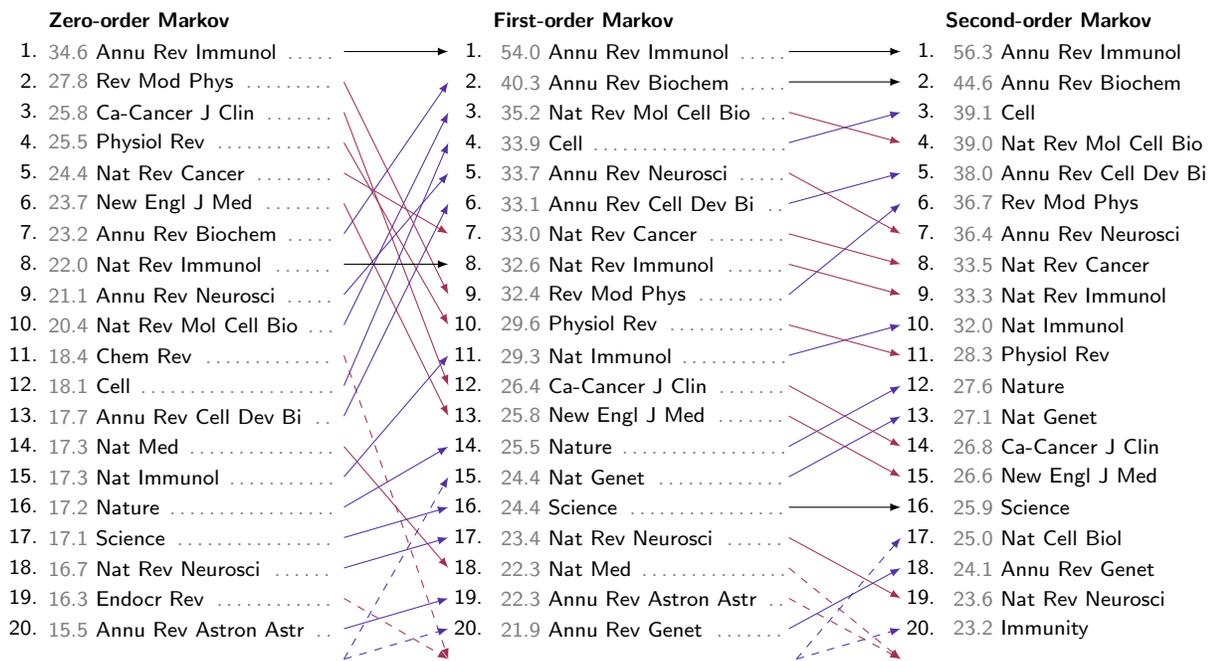

%=====================================================================

The absolute ranking scores of the top journals in Fig.~\ref{fig:ranking} show a similar increase from zero- to first- and from first- to second-order Markov dynamics. In the zero-order Markov model, the five top ranking scores are about 30 times higher than the average article, in the first-order Markov model they are about 40 times higher, and in the second-order Markov model they are about 45 times higher. Moreover, it is a trend that the higher-order Markov models give wider range of scores. This effect can be explained by their non-uniform citations values; citations from top ranked journals are worth more than citations from average ranked journals \cite{west2010big}. In the second-order Markov model with more detailed structural information and more specific re-distribution of flow value, the range of scores is even wider.

\begin{figure}%[!htbp]
\centering
\includegraphics[width=\columnwidth]{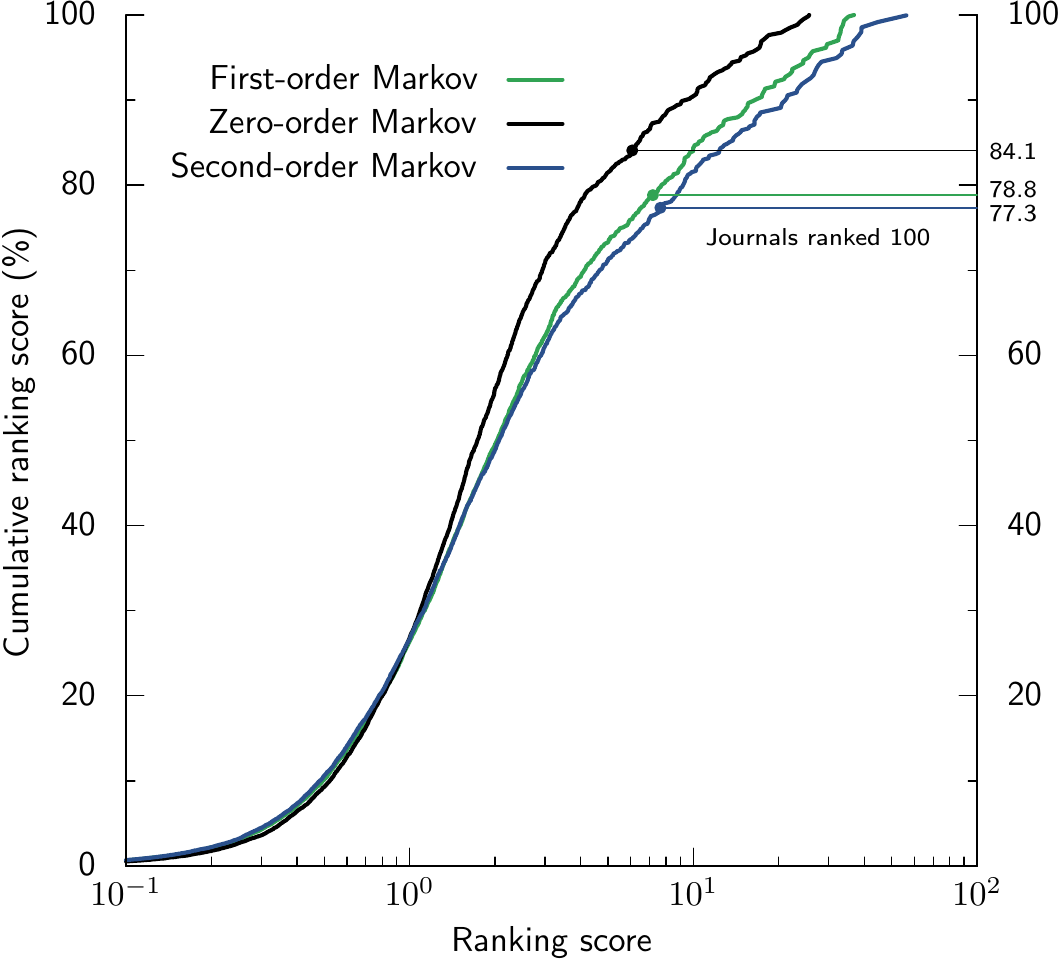}
\caption{Higher-order Markov models have wider range of ranking scores. The cumulative distribution of journal ranking scores for zero-, first-, and second-order Markov models. The points indicate the cumulative ranking scores for the journals that are ranked 100 in each ranking. \label{fig:distribution}}
\end{figure}

Figure~\ref{fig:distribution} shows the cumulative journal frequency and ranking scores. The cumulative ranking scores show that the top 100 journals in the zero-order Markov model share 15.9\% of all flow, whereas the top 100 journals in the second-order Markov model share 22.7\% of all flow. The first-order model is in between the other models with 21.2\% of all flow. Overall, the higher-order Markov models show a wide range of scores from the lowest to the highest values.

\subsection*{Comparing robustness}

\begin{figure*}%[!htbp]
\centering
\includegraphics[width=\textwidth]{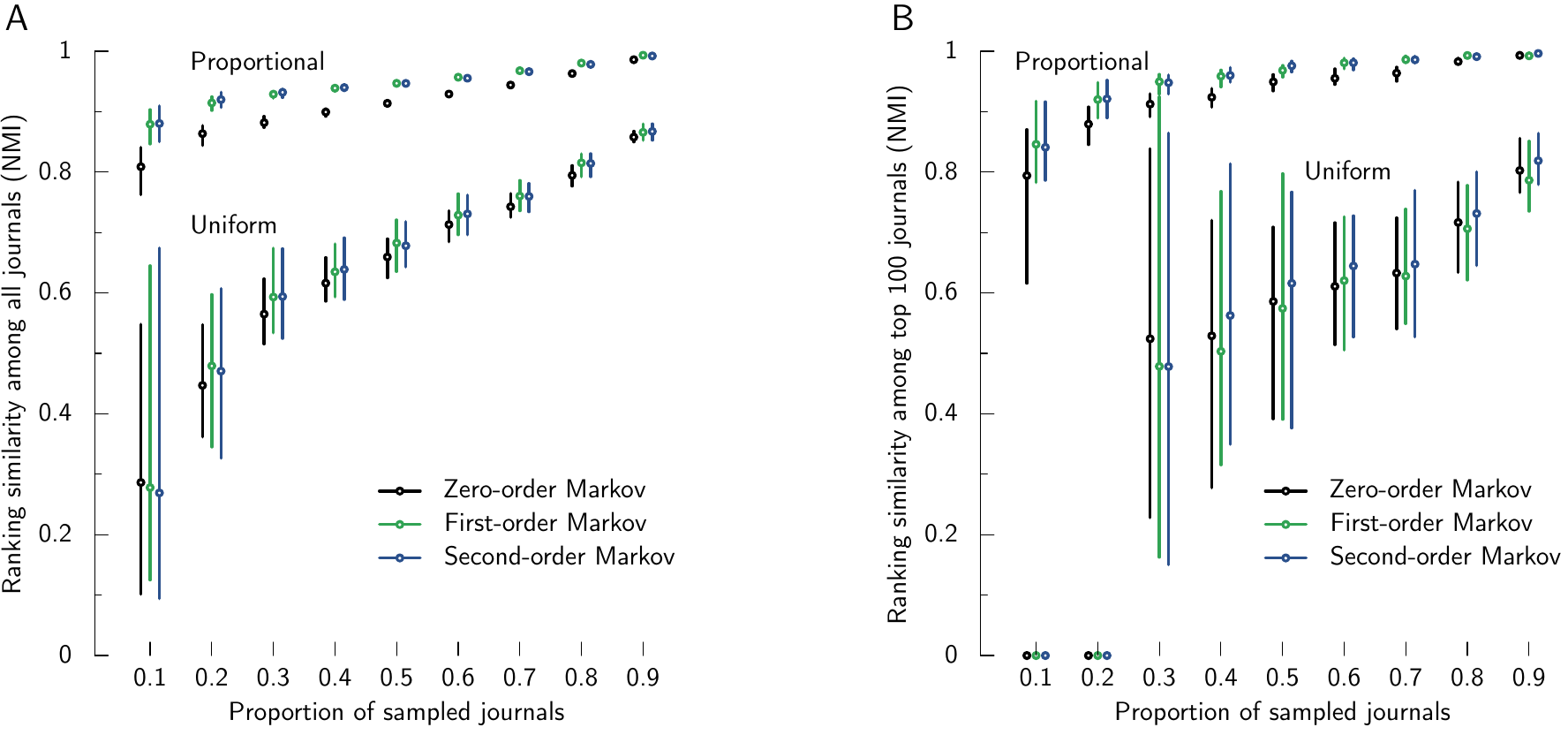}
\caption{Higher-order Markov models are more robust to the selection of journals. Ranking similarity measured by the normalized mutual information (NMI) between rankings obtained with uniform and proportional selections of included journals as a function of sample size. Ranking similarity among all journals in A and among the subset of top 100 ranked journals in B. The circles show the medians and the bars the 5th and 95th percentiles of 90 pairwise comparisons for each sample size and model. \label{fig:samplerobustness}}
\end{figure*}

A method that is good in theory is of little use if the results are not robust in practice. For journal rankings, the most crucial factor is how robust the results are to the particular selection of journals included in the analysis. For journals indexed by Thomson Reuters Web of Science, the citation data are more or less complete for the indexed journals. However, only a fraction of journals are indexed and the rankings inevitably depend on the selection. Therefore, we examined the robustness of the different models by performing analysis on random sub-samples of the set of all journals. We generated sub-samples that contained 90\%, 80\%,\ldots, 10\% of all journals by randomly including the journals. Since highly ranked journals are more likely to be included in practice, we complemented this uniform sampling with a proportional sampling in which we included journals proportional their citation counts. For each sub-sample size, we generated 10 samples and measured the ranking similarity between all pairs of rankings for each model. We used the normalized mutual information for rankings to measure the similarity \cite{lambiotte2012ranking}. This measure quantifies between 0 and 1 how much information one ranking provides about the other for journals common to both rankings. Results close to 1 mean that few journal pairs swap ranking order between rankings and indicate that the results are robust to the selection of journals.

Figure~\ref{fig:samplerobustness} shows that the ranking robustness to journal selection tends to increase with Markov order. All models become less robust with decreasing sample sizes, but the median ranking similarities among all journals are generally higher for the first- and second-order Markov models (Fig.~\ref{fig:samplerobustness}A), and among top 100 journals highest for the second-order Markov model (Fig.~\ref{fig:samplerobustness}B). For example, for the ranking similarity among top 100 ranked journals with 90\% proportionally sampled journals reported in Fig.~\ref{fig:samplerobustness}B, the second-order model is more robust than the zero- and first-order models in 89\% and 84\% of the comparisons, respectively. 

At least three factors influence the robustness: the weight of citations, how local the model is, and the range of ranking scores. In the zero-order model, all citations carry equal weight and perturbations of low-ranked journals will have as high impact as perturbations of high-impact journals. In the higher-order models, the weight of citations from low-ranked journals is reduced and has less impact on the ranking. The zero-order and also the second-order model are more local in the sense that perturbations do not propagate across the network. For the zero-order model, it is simply because only directly cited journals are affected. For the second-order model, it is because the model can capture more constrained dynamics as illustrated in Fig.~\ref{fig:schematicwalk} and demonstrated in ref.~\cite{rosvall2014memory}. Finally, the range of scores can influence the robustness in two ways. First, simply because any two journals tend to be farther apart in units of ranking score. Second, because in the normalized mutual information between rankings we make pairwise comparisons with weights proportional to the ranking scores of the journals. Since the top journals have more extreme ranking values in the higher-order models, effectively fewer journal pairs will dominate the similarity measure. All these effects together make the higher-order Markov models more robust to the selection of journals included in the analysis.

\subsection*{Cross validation}

Finally, to validate that the data are sufficient for analysis with higher-order Markov models, we conducted a cross-validation test \cite{arlot2010survey}. If the data were not sufficient, the models would overfit the data and this could lead to false conclusions. For example, the ranking scores of the higher-order Markov models could be different from the lower-order Markov models simply because of noise in sparse data.

We performed the cross-validation test by predicting movements of the random walker in the second-order Markov model in Eq.~(\ref{eq:secondorderNextstep}) with the zero-, first-, and second-order models. First, we divided all articles in 2007 into 10 random sets and generated 10 corresponding sets of trigrams, such that aggregating them all would give the complete set of trigrams. Then, in each of 10 folds, we aggregated nine sets into a training set and used the last set for validation. For each fold and each order model $o$, we measured the cross-entropy $H(p,q_\mathcal{M{\mathrm{o}}})$ of the probability distributions $p \equiv p(\vv{ij} \rightarrow \vv{jk})$ of the validation set and $q_\mathcal{M{\mathrm{o}}} \equiv q_\mathcal{M{\mathrm{o}}}(\vv{ij} \rightarrow \vv{jk})$ of the training set for the zero-, first- and second-order models $\mathcal{M_{\mathrm{o}}}$,
\begin{align}
\label{eq::cross_entropy}
H(p,q_\mathcal{M_{\mathrm{o}}}) = -\sum_{ij} p(\vv{ij}) \sum_{k} p(\vv{ij} \rightarrow \vv{jk}) \log {q_\mathcal{M_{\mathrm{o}}}(\vv{ij} \rightarrow \vv{jk})}
\end{align}
That is, we measured the cost in bits of predicting the next journal in a random walk on the validation set with transition rates obtained from the training set.

We found that navigation on the validation set costs 10.1(1) bits with the zero-order, 9.1(1) bits with the first-order, and 9.2(1) bits with the second-order Markov model fitted on the training set. Thus, the two higher-order models have a clear advantage over the zero-order model. While the two higher-order models perform similarly averaged over all journals, a journal-by-journal comparison highlights their differences. The second-order model can better predict pathways through high-impact multidisciplinary journals (see Fig.~\ref{fig:schematicwalk}C), and therefore gives a higher robustness for top 100 journals (Fig.~\ref{fig:samplerobustness}B), at an increased risk of overfitting pathways through field-specific journals with fewer citations. To quantify this effect, we derived the ratio of the posterior probabilities of the second- to the first-order model from the cross-entropy with Bayes' theorem. With uniform prior on the models $\mathcal{M_{\mathrm{2}}}$ and $\mathcal{M_{\mathrm{1}}}$, the ratio between the posterior probabilities of the two models is
\begin{equation}
\label{eq::bayes}
\frac{P(\mathcal{M_{\mathrm{2}}}|p,q_\mathcal{M_{\mathrm{2}}})}{P(\mathcal{M_{\mathrm{1}}}|p,q_\mathcal{M_{\mathrm{1}}})} = 2^{{H(p,q_\mathcal{M_{\mathrm{1}}})}{-H(p,q_\mathcal{M_{\mathrm{2}}})}}.
\end{equation}
Table \ref{tab:diff} shows that this model probability ratio is particularly high for multidisciplinary journals such as Science and Nature. Overall, the zero-order model underfits the data, the first-order model underfits multidisciplinary journals, and the second-order model has a tendency to overfit movements in field-specific journals with fewer citations, but succeeds in capturing movements in multidisciplinary journals. This result suggests that the best model is a combination of the first- and second-order Markov model.

%Top 100.

\begin{table}[htb]
\small{
\caption{Top gainers and losers among the top 100 journals in the cross-validation test. The relative difference in posterior probabilities of the second-order compared with the first-order Markov model \label{tab:diff}}
\begin{tabular}{rlr}
& \textbf{Journa}l & \textbf{Difference} \\
\hline
\rule{0em}{1.2em}1. & Nature & 162.5\% \\
2. & Science &  104.2\% \\
3. & Mat Sci Eng R & 77.1\%  \\
4. & P Natl Acad Sci USA& 67.2\%  \\
5. & Phys Rep& 14.0\%  \\
\vdots\phantom{..} & & \vdots\phantom{3\%} \\
\rule{0em}{1.2em}96. & Nat Rev Drug Discov & -30.3\% \\
97. & Nat Biotechnol & -30.4\% \\
98. & Ann Intern Med& -32.0\% \\
99. & Arch Gen Psychiat & -35.4\% \\
100. & Ca-Cancer J Clin & -35.5\% 
\end{tabular}
}
\end{table}

%%%%%%%%%%%%%%%%%%%%%%%%%%%%%%%%%%%%%%%%%%%%%%%%%%%%%%%%%%%%%%%%%%%%%%%%%%%%
%%%%%%%%%%%%%%%%%%%%%%%%%%%%%%%%%%%%%%%%%%%%%%%%%%%%%%%%%%%%%%%%%%%%%%%%%%%%
%%%%%%%%%%%%%%%%%%%%%%%%%%%%%%%%%%%%%%%%%%%%%%%%%%%%%%%%%%%%%%%%%%%%%%%%%%%%
%%%%%%%%%%%%%%%%%%%%%%%%%%%%%%%%%%%%%%%%%%%%%%%%%%%%%%%%%%%%%%%%%%%%%%%%%%%%

\section*{Conclusions}

We have shown that the robustness of flow-based rankings to the selection of included journals tends to increase with increasing Markov order. Lower-order rankings, of which impact factor is an example, depend more on the particular selection of journals because all citations carry equal weight, and because the range between the lowest and highest ranked journals is smaller than for higher-order models. Since the decision about which journals to include is difficult to make objectively and rarely made transparently, the robustness of a ranking scheme is important. Whereas our analysis indicates that higher-order models perform better, the performance gain for the second-order Markov model comes at the cost of requiring more citation data over a longer time period. While rankings can have many different objectives and be subject to various constraints that would favour other ranking schemes, if the sole objective of the ranking is to accurately capture likely pathways of researchers navigating between journals, model selection shows that using the more complex models pay off. However, the first-order Markov model underfits multidisciplinary journals and the second-order Markov model shows a tendency to overfit journals with limited data. The results suggest that an adaptive method that combines first-, second-, and even higher-order dynamics for multidisciplinary journals could further improve the ranking.

%%%%%%%%%%%%%%%%%%%%%%%%%%%%%%%%%%%%%%%%%%%%%%%%%%%%%%%%%%%%%%%%%%%%%%%%%%%%%
%%%%%%%%%%%%%%%%%%%%%%%%%%%%%%%%%%%%%%%%%%%%%%%%%%%%%%%%%%%%%%%%%%%%%%%%%%%%%
%%%%%%%%%%%%%%%%%%%%%%%%%%%%%%%%%%%%%%%%%%%%%%%%%%%%%%%%%%%%%%%%%%%%%%%%%%%%%
%%%%%%%%%%%%%%%%%%%%%%%%%%%%%%%%%%%%%%%%%%%%%%%%%%%%%%%%%%%%%%%%%%%%%%%%%%%%%

\section*{Acknowledgements}

We thank S.~Karlsson and C.~Wiklander for providing the journal citation data. M.R.\ was supported by the Swedish Research Council grant 2012-3729.

%%%%%%%%%%%%%%%%%%%%%%%%%%%%%%%%%%%%%%%%%%%%%%%%%%%%%%%%%%%%%%%%%%%%%%%%%%%%%
%%%%%%%%%%%%%%%%%%%%%%%%%%%%%%%%%%%%%%%%%%%%%%%%%%%%%%%%%%%%%%%%%%%%%%%%%%%%%
%%%%%%%%%%%%%%%%%%%%%%%%%%%%%%%%%%%%%%%%%%%%%%%%%%%%%%%%%%%%%%%%%%%%%%%%%%%%%
%%%%%%%%%%%%%%%%%%%%%%%%%%%%%%%%%%%%%%%%%%%%%%%%%%%%%%%%%%%%%%%%%%%%%%%%%%%%%

%BiBTeX style
%\bibliographystyle{plain}% plain - unsrt - abbrv
%\bibliographystyle{unsrt}
%\bibliography{2ndOrderRanking}

\end{document}